\documentclass[a4paper,pre,10pt,amsmath,amssymb,nofootinbib,showpacs,superscriptaddress,floatfix,twocolumn]{revtex4}
\usepackage[english]{babel}
\usepackage{graphicx}

\newcommand{\dfr}[2]{\frac {\displaystyle #1}{\displaystyle #2}}

\begin{document}
\title{Multi-critical dynamics of the Boson system in the vicinity of the second-order quantum phase transition}
\author{Mikhail Vasin}
\address{Physical-Technical Institute, Ural Branch of Russian Academy of Sciences, 426000 Izhevsk, Russia}
\address{High Pressure Physics Institute, Russian Academy of Sciences, Moscow, Russia}

\begin{abstract}
The non-equilibrium dynamics of the critical behaviour of the Boson system undergoing the second order quantum phase transition is discussed.
The analysis is carried out using the Keldysh technique of the non-equilibrium dynamics description.
The critical behaviour close to the quantum critical point has been shown to be multi-critical. Crossover among three different critical regimes (modes) is possible: adiabatic quantum mode (AQM), dissipative classical mode (classical critical dynamics mode (CCDM) and dissipative quantum critical mode (QCDM). As a result, the observed critical behavior can essentially depend on the conditions of approaching the critical point.
\end{abstract}

\maketitle

Recently there has been considerable interest in the experimental and theoretical studies of the quantum phase transition dynamics. This interest is quite natural as quantum phase transitions are essentially dynamic \cite{Hertz}. In them the time acts as an equal additional space dimension. However, usually one considers only a simplified case assuming that in the vicinity of the critical point it is possible to distinguish two regimes: in one of them the energy of thermal fluctuations exceeds the energy of quantum fluctuations,  $k_{B}T\gg \hbar \omega_{\gamma }$ ($\omega_{\gamma }$ is the quantity reciprocal to the time of the system relaxation, $\tau_{\gamma}$), the critical mode being described by classical dynamics; in the other one the energy of thermal fluctuations becomes less than the energy of quantum fluctuations, $k_{B}T\ll \hbar \omega_{\gamma }$, the system being described by quantum mechanics  \cite{Hertz,Sachdev}.
This description is not complete since it does not take into account the effect of dissipation in the quantum fluctuation regime, though it is well known that dissipation drastically change the critical properties \cite{L,W1,W2}.
It is clear that the system turns from the mode of dissipative dynamics of thermal fluctuations into the adiabatic mode of purely quantum fluctuations, then there should exist some intermediate dissipative mode of quantum fluctuations. The crossover between these critical modes has not been theoretically studied so far. Besides, nobody has considered the question on the influence of the ratio of the maximum observation time, $t_{max}$, which determines the least allowed value $\omega _0\propto 1/t_{max}$ of the scale of the quantum fluctuations energy, on the value $\tau_{\gamma }$.
It will be shown below that one should not neglect it. In addition, within a unified approach based on the Keldysh technique of non-equilibrium dynamics description, the crossover among all three critical modes in the vicinity of the quantum critical point will be described.

To describe quantum critical dynamics theoretically, it is most convenient to use the Keldysh technique initially formulated for quantum systems. Let the system of our interest be the Boson system, whose state is described with the real field of the order parameter $\phi $, and the potential energy is determined by the functional
$U(\phi )$, e.g. $U\propto \phi^4$. Let us assume that $\hbar =1$ and $k_{B}=1$. In the static, to say more correctly, in the stationary, not quantum case the physics of the system is determined by the static sum:
\begin{gather*}
Z=N\int \mathfrak{D}\phi\exp \left[-S(\phi)\right],
\end{gather*}
where $\int \mathfrak{D}\phi$ denotes the functional $\phi $-field integration, $S$ is the action that in the general form is as follows:
\begin{gather*}
S(\phi)=\dfr 1{T}\int dk \left( \phi^{\dag} G^{-1}\phi + U(\phi)\right),\\
G^{-1}=\varepsilon_k=k^2+\Delta,
\end{gather*}
where $T$ is themperature of the system, $U$ is the potential energy depending on the order parameter, $\Delta $ is the governing parameter, that tends to zero at the critical point.

There are different existing methods used for the transition from equilibrium statics to non-equilibrium dynamics. Note, that all of them result in doubling the fields   describing the system, suggest the interaction of the system with the thermostat and are essentially equivalent to each other. As it has been mentioned above, we are going to use the Keldysh technique, which seems most convenient. In this case the role of the statistic sum is played by the functional path integral that after Wick rotation is as follows \cite{Sachdev}:
\begin{gather*}
\displaystyle Z=N\int \mathfrak{D}\phi^{cl} \mathfrak{D}\phi^{q}\exp \left[-S(\phi^{cl}, \phi^{q})\right],\\
\displaystyle S(\phi^{cl}, \phi^{q})=\int d\omega dk \left( \bar\phi^{\dag} \hat G^{-1}\bar\phi + U(\phi^{cl}+\phi^q)\right. \\
\displaystyle \left.-U(\phi^{cl}-\phi^q)\right),\quad \bar\phi =\{\phi^{q},\,\phi^{cl}\},
\end{gather*}
where $\phi^{q}$ and $\phi^{cl}$ are pair of fields called ``quantum'' and ``classical'' respectively. In the case of the Boson system the matrix of the inverse correlation functions is the following \cite{Sachdev}:
\begin{gather}
\hat G^{-1}=\left[ \begin{array}{cc}\displaystyle 0 & \omega^2+\varepsilon_k+i\gamma\omega \\ \displaystyle \omega^2+\varepsilon_k-i\gamma\omega & 2\gamma \omega \coth(\omega/T) \end{array}\right],
\label{R1}
\end{gather}
where $\gamma $ is the kinetic coefficient, and the function $2\omega \coth(\omega/T)$ is the function of the density of states of the ideal Boson gas. The
advancing, retard and Keldysh parts of both the correlation functions matrix and the inverse matrix are connected by the relation known as the fluctuation-dissipation theorem (FDT):
\begin{gather*}
\displaystyle [\hat G]^{K}=i\coth(\omega/T)\left[[\hat G]^{A}-[\hat G]^{R} \right]\\ =2\coth(\omega/T)\,\mbox{Im} \left([\hat G]^{R}\right),\\
\displaystyle [\hat G^{-1}]^{K}=i\coth(\omega/T)\left[[\hat G^{-1}]^{A}-[\hat G^{-1}]^{R} \right]\\=-2\coth(\omega/T)\,\mbox{Im} \left([\hat G^{-1}]^{R}\right).
\end{gather*}

The expressions given above allow us to describe the critical dynamics of the system theortically in the vicinity of the critical point. They are general and allow the system to be described both within the classical, $T\gg \omega$,  and the quantum, $\omega \gg T$, limits.

\begin{figure}[h]
\centering
   \includegraphics[scale=0.6]{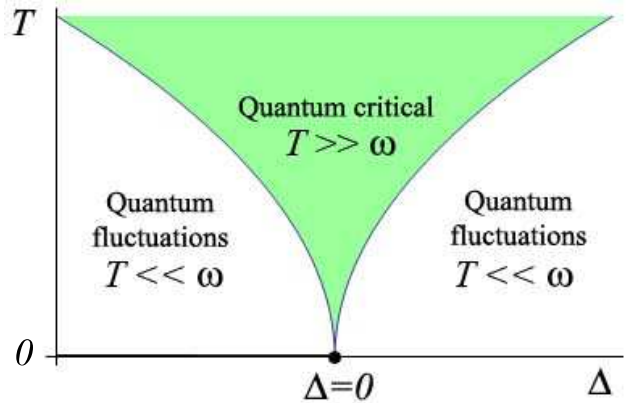}
   \caption{($T=0, \Delta =0$) is the quantum critical point. The lines indicate the boundaries of the quantum critical region where the leading critical singularities can be observed; these crossover lines are given by $T\propto \Delta^{\nu z}$ ($z$ and $\nu $ are the critical exponents). Whether it is strange that the mode with the thermal fluctuations dominance ($T\gg \omega$) is called as a quantum one?}
   \label{F1}
\end{figure}

\begin{figure}[h]
\centering
   \includegraphics[scale=0.45]{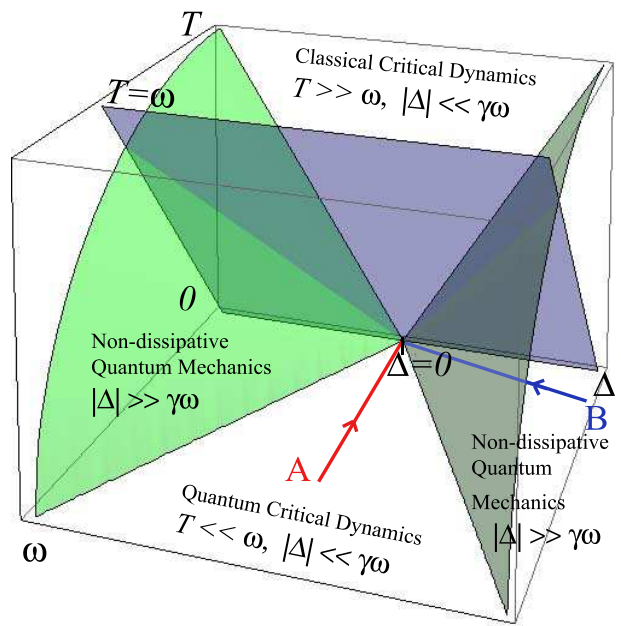}
   \caption{The green colour denotes the conventional surface $\omega^2+T^2=|\Delta|^{2\nu}$ showing the location of the crossover field between the dissipative and adiabatic fluctuation modes.}
   \label{F2}
\end{figure}

Usually to illustrate the crossover between different critical modes in the vicinity of the quantum critical point, the plots of the temperature versus the governing parameter are referred to, given in Fig.\,\ref{F1}.
But this picture is not quite correct.
To show all possible critical modes, it is necessary to expand the coordination space  by adding $\omega$, corresponding to the energy of quantum fluctuations. Fig.\,\ref{F1} considers the limiting case $\omega =0$, that, strictly speaking, is experimentally unattainable. Since any experiment on a specific system is carried out for a finite time, then in the theoretical description of this system one should take into account the relationship of the maximum time of observation with the time of its relaxation (or the time of coherence). As it is shown in Fig.\,\ref{F2}, first of all, the three-dimensional parametric space can be divided into two regions: $T\gg \omega$ and $T\ll \omega$, which, as it was mentioned above, correspond to the fields of the classical and quantum description.

Let us consider the first region, where thermal fluctuations dominate, $\omega \ll T$. Note that the plane $\omega =0$ is entirely located in this region.
The critical dynamics of the system is determined by the Keldysh element of the matrix of Green functions, $[\hat G^{-1}]^K=2\gamma \omega \coth(\omega/T)$.
Within $T\gg \omega$ this function tends to
\begin{gather*}
\lim\limits_{T\gg \omega } [\hat G^{-1}]^K \approx 2\gamma T.
\end{gather*}
Note that in this case the effect of the thermostat on the system (the action of the statistical ensemble on its own element) corresponds to the influence of the external ``white'' noise. The fluctuations with the smallest wave vectors and energies ($k\to 0$, $\omega\to 0$) are considered to be relevant (significant) in the vicinity of the critical point, hence only the terms with the lowest degrees $k$ and $\omega $ are retained in the expressions.
As a result, in the fluctuation field the system is described dy the standard classical non-equilibrium dynamics:
\begin{gather*}
\hat G^{-1}=\left[ \begin{array}{cc}0 & \varepsilon_k+i\gamma\omega \\ \varepsilon_k-i\gamma\omega & 2\gamma T \end{array}\right].
\end{gather*}
satisfying the standard form of FDT:
\begin{gather*}
[\hat G^{-1}]^{K}=\dfr{T}{\omega }\mbox{Im} \left([\hat G^{-1}]^{R}\right).
\end{gather*}
The dispersion relationship in this case is: $\omega \propto k^2$,
whence it follows that the dynamic critical exponent of the theory (scaling dimension) in the first approximation will be: $z=2$.
The dimension of the system is: $D=d+z=d+2$, but due to the ``white'' noise presence the effective dimension of the system is:
$D_{ef}=D-2=d$ \cite{Parisi}.
Naturally, it results in the coincidence of the critical dimensions of the dynamic and static theories, the critical behaviour of the system being described by the classical critical dynamics of the  $d$-dimensional system. Let us refer to this mode as the mode of the classical critical dynamics (CCDM).

With $\omega \ll T$ the case is possible when the time of the system relaxation turns out to be much shorter than the characteristic period of thermal fluctuations, $\gamma T \ll \Delta $. Then the dissipation may by neglected ($\gamma \to 0$), thus:
\begin{gather*}
\hat G^{-1}\approx \left[ \begin{array}{cc}0 & \omega^2+\varepsilon_k\\  \omega^2+\varepsilon_k & 0\end{array}\right],
\end{gather*}
the dispersion relationship takes the form $\omega \propto k$, and az a result, $z=1$. In this region the critical behaviour is described as the critical behaviour of the static system with the dimension $d+1$.

Now let us consider the field in which quantum fluctuations dominate, $\omega\gg T$, in which the system ``does not know'' that it has got temperature. In this case (\ref{R1}) will be follows:
\begin{gather*}
\hat G^{-1}\approx \left[ \begin{array}{cc}0 & \varepsilon_k+i\gamma\omega \\ \varepsilon_k-i\gamma\omega & 2\gamma |\omega | \end{array}\right].
\end{gather*}
In spite of the absence of thermal fluctuation in the quantum case FDT still exists and has the following form:
\begin{gather*}
[\hat G^{-1}]^{K}=-2\,\mbox{sign}(\omega )\mbox{Im} \left([\hat G^{-1}]^{R}\right),
\end{gather*}
and the action of the statistic ensemble on the system does not depend on the temperature.
Note, that close to the phase transition ($\Delta \approx 0$), when $\varepsilon _k\to 0$, we get $G^K(\omega )={2}/{\gamma |\omega |}$.
This is the so called $1/f$-noise (Flicker noise), whose intensity does not depend on the temperature. 
The latter changes the critical properties of the system significantly.  As in the case of classical critical dynamics the dimension in this case is $D=d+2$. However, the $1/f$ noise in contrast to the ``white''-noise, does not decrease the effective dimension  \cite{Vasin}, therefore the dimension of the dissipative quantum system is less by 2 than its static dimension, $D_{ef}=d+2$. The disagreement of the static and dynamic theories is accounted for by the fact that in the quantum case there is no statistic limit, and the only correct results are those of the dynamic theory. This dynamic mode can be referred to as the mode of the quantum critical dynamics (QCDM).

In the quantum region as well as in the region corresponding to classical dynamics, when the time of coherence of the system appears much shorter than the inverse frequency of quantum fluctuations, $\gamma \omega  \ll |\Delta |$, the dynamics of the system changes into an adiabatic mode, in which the dissipation can be neglected:
\begin{gather*}
\hat G^{-1}\approx \left[ \begin{array}{cc}0 & \omega^2+\varepsilon_k\\  \omega^2+\varepsilon_k & 0\end{array}\right].
\end{gather*}
By analogy the dissipation relation acquires the following form $\omega \propto k$, and, as a result, $z=1$. Thus, in this field the critical behaviour is described as the critical behaviour of the static system with the dimension $d+1$. It is easy to see that this is the field in which the Matsubara formalism works. Its joining up with the classical adiabatic field gives a common region of adiabatic dynamics which can be referred to as the dynamics in the adiabatic quantum mechanical mode (AQM).

Let us consider separately the field of ``crossing'' of all modes, which is in the vicinity of the plane $\omega =T$. Here the thermal and quantum fluctuations are equal, thus this field is the field of crossover between classical and quantum dynamic modes. Since the critical dynamics in these modes is different, then the exact determinationof the critical exponents in the fluctuation (dissipative) field is impossible. However, in the field with the adiabatic mode of dynamics the critical exponents are well determined for they do not depend on the relation of thermal and quantum fluctuations energies.

From the arguments mentioned above one can distinguish three fields in the parametric space that differ by the modes of critical dynamics  (Fig.\,\ref{F2}). The critical exponents in these fields are differ.  For a specification let us consider, for example, the critical (fluctuation) field in the vicinity of the quantum phase transition of the second order ($\Delta =0$, $T=0$) in the one-dimensional Ginsburg--Landau model. In this case the field of the order parameter is a real field $\phi $, and $U(\phi )=\phi^4$. Let us consider the field with different modes of critical dynamics: 1) In the field in which thermal fluctuations dominate, that is in CCDM, the space dimension of the system is below than the low critical dimension of the given model,  $d=1<d_c^{-}=2$. It means that formally the phase transition is not possible since any ordering in a one-dimensional case is destroyed by the thermal fluctuations according to the Mermin--Wagner  theorem. However, to realize such a mode is practically impossible. The most likely is the realization of one mode from two others;  2) In the non-dissipative AQM the coherence time is negligibly less than the time of observing the system, thus the relaxation processes can be neglected. In this case, as it was shown above, the critical behaviour of the dynamics system is analogous to the critical behaviour of the static system of the $d+1=2$ dimension. The critical exponents of this system are well known. So, for example, the critical exponent $\nu $, determining the degree of divergence of the correlation radius, $r_c\propto |\Delta |^{-\nu}$, is determined exactly as $\nu = 1$. The coherence time here depends on $\Delta $ in the same way as the correlation radius: $\tau \propto r_{c}\propto |\Delta |^{-1} $. We can denote the boundary of AQM with the help of this dependence (Fig.\,\ref{F2}). Thus, different modes are separated by this surface and the $T=\omega $ surface.  Note that this separation is not sharp, the planes only show the location of blurred regions of the crossover between the modes. Besides, it should be noted that the upper critical dimension of the system under consideration is equal to $d_c^+=3$, which means that the critical exponents of the three-dimensional system being within the considered mode, is determined well with the help of the mean field theory; 3) The most probable mode in experimental studies seems to be the dissipative QCDM. This mode of critical dynamics of the one-dimensional Ginsburg--Landau model was analysed in \cite{Vasin}. In the one-loop approximation the critical exponent has found $\nu \approx 0.6$, and the relation for the coherence time was obtained, $\tau_{\gamma } \propto -|\Delta|^{-1}\ln|\Delta |$.

Thus, the functional technique of theoretical description of non-equilibrium dynamics allows us to describe the entire spectrum of critical modes in the vicinity of quantum phase transition within a single formalism.  The main conclusion of this work is that the critical behaviour in the vicinity of the quantum critical point is multi-critical. As a result, the observed critical behaviour depends on the conditions of approaching the critical point  to a great extent, i.e. on the ``way'' the experimenter approaches it. If one approaches the critical point quickly ($T=0$, $\Delta =0$) and then waits long (trajectory A in Fig.\,\ref{F2}), then the critical exponents will be tend to the values predicted by the quantum critical dynamics.  If at first the system is cooled down to  $T=0$, then the governing parameter begins to approach the critical point, $\Delta \to 0$ (trajectory B in Fig.\,\ref{F2}), the measured critical exponents will correspond to the adiabatic mode .

This work was partly supported by the RFBR grants No. 13-02-91177 and No. 13-02-00579.


\begin{thebibliography}{99}

\bibitem{Hertz} Hertz J. A. Quantum critical phenomenal, Phys.Rev.B {\bf 14,} 1165--1184 (1976);
\bibitem{Sachdev} Sachdev S, Quantum Phase Transitions, Cambridge University Press, New York,
ISBN 978-0-521-51468-2, 501 p., 2011;
\bibitem{L} Leggett A.J, Chakravarty S, Dorsey AT, Fisher MPA, Garg A, and Zwerger
W. Dynamics of the dissipative two-state system. Rev. Mod. Phys.
{\bf 59,} 1--85 (1987); Weiss U. Quantum Dissipative Systems. World Scientific,
Singapore, 1999;
\bibitem{W1} Werner P, V\"{e}olker K, Troyer M, and Chakravarty S. Phase Diagram and
Critical Exponents of a Dissipative Ising Spin Chain in a Transverse
Magnetic Field. Phys. Rev. Lett. {\bf 94,} 047201--047204 (2005);
\bibitem{W2} P. Werner, M. Troyer, and S. Sachdev. Quantum Spin Chains with Site
Dissipation. J. Phys. Soc. Jpn. Suppl. {\bf 74,} 67--70 (2005).
\bibitem{Parisi} G. Parisi, N. Sourlas, Phys.\,Rev.\,Lett. {\bf 43,} 744 (1979);
\bibitem{Vasin} Vasin M. G. Quantum critical dynamics of the boson system in the Ginsburg-Landau model, arXiv:1401.1304.


\end{thebibliography}
\end{document}